\documentstyle[12pt,fullpage,epsfig]{article}
\def\be{\begin{equation}}
\def\ee{\end{equation}}
\def\beq{\begin{eqnarray}}
\def\eeq{\end{eqnarray}}
\def\lsim{\:\raisebox{-0.5ex}{$\stackrel{\textstyle<}{\sim}$}\:}
\def\gsim{\:\raisebox{-0.5ex}{$\stackrel{\textstyle>}{\sim}$}\:} 
\begin{document}
\begin{flushright}
RAL-TR-1999-043 \\
TSL/ISV-99-0217 \\
TIFR/TH/99-49 \\
September 1999
\end{flushright}
\bigskip
\begin{center}
{\Large{\bf Detecting heavy charged Higgs bosons at the LHC \\[0.5cm]
with triple $b$-tagging}} \\[1cm]
{\large S. Moretti$^{1,2}$ and D.P. Roy$^3$} \\[1cm]
$^1${\sl Rutherford Appleton Laboratory, Chilton, Didcot, Oxon OX11
0QX, UK} \\
$^2${\sl Dept. of Radiation Sciences, Uppsala University,
P.O.~Box~535,~75121~Uppsala,~Sweden} \\
$^3${\sl Tata Institute of Fundamental Research, Mumbai - 400 005,
India} \\[1.5cm]
{\bf Abstract} 
\end{center}
\bigskip

We investigate the charged Higgs boson signal at the LHC using its
dominant production and decay modes with triple $b$-tagging, i.e. $tH^-
\rightarrow t\bar tb \rightarrow b\bar bb W^+ W^-$, followed by
leptonic decay of one $W$ and hadronic decay of the other.  We
consider the continuum background from the associated production of
$t\bar t$ with a $b$- or a light quark or gluon jet, which can be
mis-tagged as $b$-jet.  We reconstruct the top quark masses to identify
the 3rd $b$-jet accompanying the $t\bar t$ pair, and use its $p_T$
distribution to distinguish the signal from the background.  Combining
this with the reconstruction of the $H^\pm$ mass gives a viable
signature over two interesting regions of the parameter space --
i.e. $\tan\beta \sim 1$ and $\sim m_t/m_b$.

\newpage

The Minimal Supersymmetric Standard Model (MSSM) contains two complex
Higgs doublets, $\phi_1$ and $\phi_2$, corresponding to eight scalar
states.  Three of these are absorbed as Goldstone bosons leaving five
physical states -- the two neutral scalars $(h^0,H^0)$, a
pseudo-scalar $(A^0)$ and a pair of charged Higgs bosons $(H^\pm)$.
All the tree-level masses and couplings of these particles are given
in terms of two parameters, $M_{H^\pm}$ and $\tan\beta$, the latter
representing the ratio of the vacuum expectation values of 
$\phi_1$ and $\phi_2$ [1].
While any one of the above neutral Higgs bosons may be hard to
distinguish from that of the Standard Model, the $H^\pm$ carries a
distinctive signature of the Supersymmetric
(SUSY) Higgs sector.  Moreover the
couplings of the $H^\pm$ are uniquely related to $\tan\beta$, since
the physical charged Higgs boson corresponds to the combination 
\be
H^\pm = -\phi^\pm_1 \sin\beta + \phi^\pm_2 \cos \beta.
\label{one}
\ee
Therefore the detection of $H^\pm$ and measurement of its mass and
couplings are expected to play a very important role in probing the
SUSY Higgs sector. 

Unfortunately it is very hard to extend the $H^\pm$ search beyond the
top quark mass at the Large Hadron Collider
(LHC), because in this case the combination of
dominant production and decay channels, $tH^- \rightarrow t\bar t b$,
suffers from a large QCD background.  The viability of a $H^\pm$ signal
in this channel had been investigated in [2,3] assuming triple
$b$-tagging.  Recently it was shown that with four $b$-tags one can
get a better signal/background ratio, but at the cost of a smaller
signal size [4].  Similar conclusions were also found for the $H^\pm$
signal in its $\tau$ decay channel [5].  The charged Higgs boson
signal at the LHC has also been investigated recently in subdominant
production channels, $H^\pm W^\mp$ [6] and $H^\pm H^\mp$ [7], as well
as the subdominant decay mode $H^\pm \rightarrow W^\pm h^0$ [8].
But it turns out to be at best marginal in each of these cases.

It is clear from the above discussion that the largest size of the
$H^\pm$ signal is expected to come from its dominant production and
decay channels with triple $b$-tagging.  
The purpose of
this paper is to reinvestigate the $H^\pm$ signal in this channel in
the light of the theoretical and experimental developments since the
last analyses [2,3].  Several distinctions of the present study in
comparison with those earlier ones are worth mentioning here. 

\begin{enumerate}
\item[{i)}] The signal cross-section was calculated in [2] and [3]
using the $2 \rightarrow 2$ and $2 \rightarrow 3$ processes
respectively, i.e. 
\be
gb \rightarrow tH^- + {\rm h.c.},
\label{two}
\ee
and
\be
gg \rightarrow t\bar b H^- + {\rm h.c.},
\label{three}
\ee
followed by the $H^- \rightarrow \bar tb$ decay. Here  we shall instead
combine the
two cross-sections and subtract out the overlapping piece to avoid
double counting, as suggested in [9,10].
\item[{ii)}] We shall use the $p_T$ distribution of the 3rd $b$-tagged
jet, accompanying the $t\bar t$ pair, for a better separation of the
signal from the background.
\item[{iii)}] The actual value of top quark mass (175 GeV) will be
used here instead of the illustrative values used in [2,3].
\item[{iv)}] Besides we shall be using current estimates of the
$b$-tagging efficiency and rapidity coverage for the LHC [11] along with
more recent structure functions [12,13]. 
\end{enumerate}

The cross-section for the $2 \rightarrow 2$ process (2) is simple to
calculate, while analytic expressions for the $2 \rightarrow 3$
processes (3) can be found in [4].  The resulting signal
cross-sections shall be obtained by convoluting these partonic
cross-sections with the MRS-LO(05A) parton densities [12].  We have also
checked that essentially identical results are obtained with the
CTEQ4L parton densities [13].  It may be noted here that both these
cross-sections are controlled by the Yukawa coupling of the
$tbH$ vertex, 
\be
{g \over \sqrt{2} M_W} H^+ \left[\cot\beta m_t \bar t b_L + \tan\beta
m_b \bar t b_R\right] + {\rm h.c.}
\label{four}
\ee
Consequently one gets fairly large values of the signal cross-section
at the two ends of the MSSM allowed region, 
\be
\tan\beta \sim 1 ~{\rm and}~ \tan\beta \sim m_t/m_b,
\label{five}
\ee
with a pronounced minimum at $\tan\beta = \sqrt{m_t/m_b}$. 

The question of overlap between the two $H^\pm$ production processes
(2) and (3) has been recently discussed in [9,10].  The $b$-quark in
(2) comes from a gluon in the proton beam splitting into a collinear
$b\bar b$ pair, resulting in a large factor of $\alpha_S \log(Q/m_b)$,
where the factorisation scale is
\be
Q \simeq m_t + M_{H^\pm}.
\label{six}
\ee
This factor is then resummed to all orders, $\alpha_S^n \log
^n(Q/m_b)$, in evaluating the phenomenological $b$-quark structure function
[14,15].  The 1st order contribution to the structure function is
given by the perturbative solution to the DGLAP equation,
\be
b'(x,Q) = {\alpha_S \over \pi} \log \left({Q \over m_b}\right)
\int^1_x {dy \over y} P_{gb} \left({x \over y}\right) g(y,Q),
\label{seven}
\ee
where $P_{gb} (z) = (z^2 + (1-z)^2)/2$ is the gluon splitting
function.  The resulting contribution to $gb \rightarrow t H^-$ is
already accounted for by $gg \rightarrow t \bar b H^-$ in the
collinear limit.  Thus while combining (2) and (3), the above
contribution should be subtracted from the former to avoid double
counting. 

\begin{figure}[h!]
\begin{center}
{\epsfig{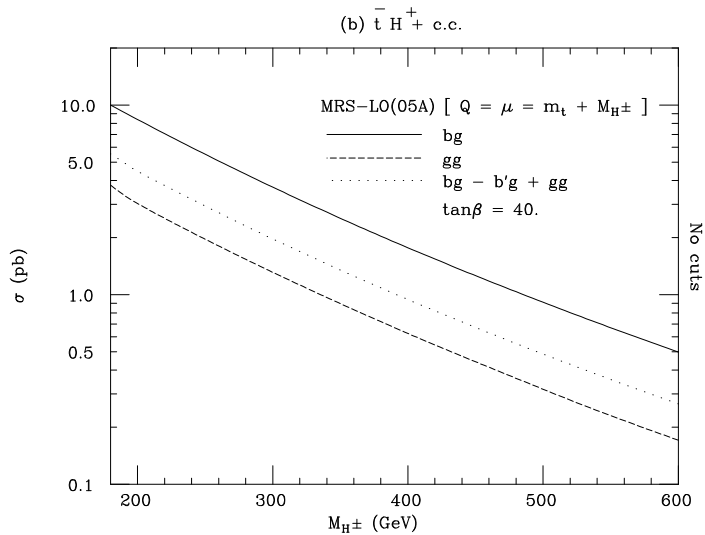}}
\caption{\small Cross section of the $2\to2$ process
$gb\to t H^-$  (2), of the $2\to3$ one
$gg\to t\bar b H^-$ (3) and of their sum after the 
subtraction of the $g b'$ contribution, see eq.~(7), 
for $\tan\beta=40$ (including the charged conjugated
final states). 
The PDF set used was MRS-LO(05A)
with renormalisation and factorisation scales set equal 
to $m_t+M_{H^\pm}$.  
Normalisation is to the total cross sections without any branching ratios.}
\label{fig:sub}
\end{center}
\end{figure}

Fig. 1 shows the cross-sections for (2) and (3) at the LHC energy
$(14~{\rm TeV})$ against the $H^\pm$ mass at $\tan\beta = 40$, using 
$m_b = 4.5 ~{\rm GeV}$.  It also shows their combined value, after
subtracting out the $gb'$ contribution from the former.  While the
cross-section for (2) is 2--3 times larger than that for (3), the bulk
of the former is accounted for by the $gb'$ contribution.  Hence the
combined cross-section is larger than that of (3) by only a factor of
about 1.6.  We also have checked that detection efficiencies for the
processes (2) and (3) are very similar, since the extra $b$-jet in the
latter case is relatively soft (missing the $p_T > 30$ selection cut
discussed below over 70\% of the time).  We shall therefore simply
multiply the cross-section for the $2 \rightarrow 3$ process (3) by
the above mentioned factor of 1.6 in presenting the signal
cross-sections. 

It should also be mentioned here 
that the electroweak loop corrections to the $tbH$
vertex (4) have been estimated to give up to 20\% reduction in the
signal cross-section depending on $M_{H^\pm}$ and $\tan\beta$ [16].  The
corresponding QCD corrections are expected to be larger, but not yet
available.  Note that higher-order QCD effects in the $H^- \rightarrow \bar
tb$ decay are easily accounted for by using the running value of the $b$
mass, $m_b (M_{H^\pm})$, in the $tbH$ coupling.  We shall therefore use it
in estimating the $H^- \rightarrow \bar t b$  decay rate.
 (Of course this has no significant impact on the signal since
this branching fraction amounts to $\gsim 80\%$ over most of the parameter 
space
of our interest.)  The effects of SUSY QCD corrections may be larger,
depending on the SUSY parameters [17].  We shall neglect this by
assuming a large SUSY mass scale $\sim 1$ TeV.

The final state resulting from the above $2 \rightarrow 2 ~(2
\rightarrow 3)$ signal process is
\be
t H^-(\bar b) \rightarrow t\bar t b(\bar b) \rightarrow b\bar b b(\bar
b) W^+W^-.
\label{eight}
\ee
We shall require leptonic decay of one $W$ and hadronic decay of the
other, resulting in a final state of
\be
b\bar bb (\bar b) \ell \nu q \bar q.
\label{nine}
\ee
The hard lepton $(e,\mu)$ will be required for triggering and
suppression of multi-jet background, while the presence of only one
$\nu$ will enable us to do mass reconstruction.  As mentioned earlier,
the extra $b$-quark coming from the $2 \rightarrow 3$ process (3) is
expected to be too soft to pass our selection cuts or be tagged with a
reasonable efficiency.  We shall therefore require a minimum of 3
$b$-tagged and 2 untagged jets along with a lepton and a missing-$p_T$
$({p\!\!/}_T)$.

We shall consider the background to the final state (9) coming from 
\be
gb \rightarrow t\bar t b \rightarrow b\bar b b W^+ W^-,
\label{ten}
\ee
\be
gg,q\bar q \rightarrow t\bar t g^\star \rightarrow b\bar b b \bar b W^+
W^-, 
\label{eleven}
\ee
along with those from 
\beq
gg,q\bar q &\rightarrow& t\bar t g \rightarrow b\bar b g W^+W^-, \nonumber
\\[2mm] g q &\rightarrow& t\bar t q \rightarrow b\bar b q W^+W^-,
\label{twelve}
\eeq
where the gluon or light quark jet $(j)$ is mis-tagged as a $b$-jet.  In
fact (12) will turn out to be the largest background.  The
cross-sections for processes (10)--(12) are computed using MadGraph and HELAS 
[18,19]. 

Our analysis is based on simply a parton level Monte Carlo program.
However we have tried to simulate detector resolution by a Gaussian
smearing of all jet momenta with [2]
\be
\left(\sigma(p_T)/p_T\right)^2 = (0.6/\sqrt{p_T})^2 + (0.04)^2,
\label{thirteen}
\ee
and the lepton momentum with
\be
\left(\sigma(p_T)/p_T\right)^2 = (0.12/\sqrt{p_T})^2 + (0.01)^2.
\label{fourteen}
\ee
The ${p\!\!/}_T$ is obtained by vector addition of all the $p_T$'s
after resolution smearing. 

As a basic set of selection cuts we require
\be
p_T > 30 ~{\rm GeV} ~~ {\rm and} ~~ |\eta| < 2.5
\label{fifteen}
\ee
for all the jets and the lepton, where $\eta$ denotes pseudorapidity
and the $p_T$-cut is applied to the ${p\!\!/}_T$ as well.  We also
require a minimum separation of ($\phi$ is the azimuthal angle)
\be
\Delta R = \left[(\Delta \phi)^2 + (\Delta \eta)^2\right]^{1/2} > 0.4
\label{sixteen}
\ee
between the lepton and the jets as well as each pair of jets.

To improve the signal/background ratio and to estimate the $H^\pm$
mass we follow a strategy similar to that in [2], except for the step
(e) below, which is new. 

\begin{enumerate}
\item[{(a)}] The invariant mass of two untagged jets is required be
consistent with $M_W \pm 15$ GeV.
\item[{(b)}] The neutrino momentum is reconstructed by equating
$p^T_\nu$ with ${p\!\!/}_T$ and fixing $p^L_\nu$ within a quadratic
ambiguity via $m(\ell\nu) = M_W$.
\item[{(c)}] The invariant mass of the above untagged jet pair with
one of the 3 $b$-tagged jets is required to be consistent with $m_t
\pm 25$ GeV.  If several $b$-tagged jets satisfy this, the one giving
the best agreement with $m_t$ is selected. 
\item[{(d)}] The invariant mass of the $\ell$ and $\nu$ with one of
the 2 remaining $b$-jets is required to be consistent with $m_t \pm
25$ GeV.  In case of several combinations satisfying this, the one
giving best agreement with $m_t$ is selected along with the
corresponding $b$-jet and $p^L_\nu$.
\item[{(e)}] The remaining (3rd) $b$-jet is the one accompanying the
$t\bar t$ pair in the signal (8) or in the backgrounds (10)--(12)\footnote{In
case of a 4th $b$-jet surviving the $p_T > 30$ GeV cut the harder of
the two $b$-jets accompanying the 
$t\bar t$ pair is selected.}.  For the signal it mainly comes from the
$H^- \rightarrow \bar t b$ decay and is therefore quite hard, while it
is expected to be very soft for the background processes.  Hence the
$p_T$-distribution of this $b$-jet shall be used to improve the
signal/background ratio. 
\item[{(f)}] Finally we combine each of the top
(anti)quarks in the
reconstructed $t\bar t$ pair with the 3rd $b$-jet.  Thus we obtain 2
entries per event in the $M_{bt}$ invariant mass
plot, one of which would correspond
to the $H^\pm$ mass peak for the signal.
\end{enumerate}

\begin{figure}[h!]
\begin{center}
{\epsfig{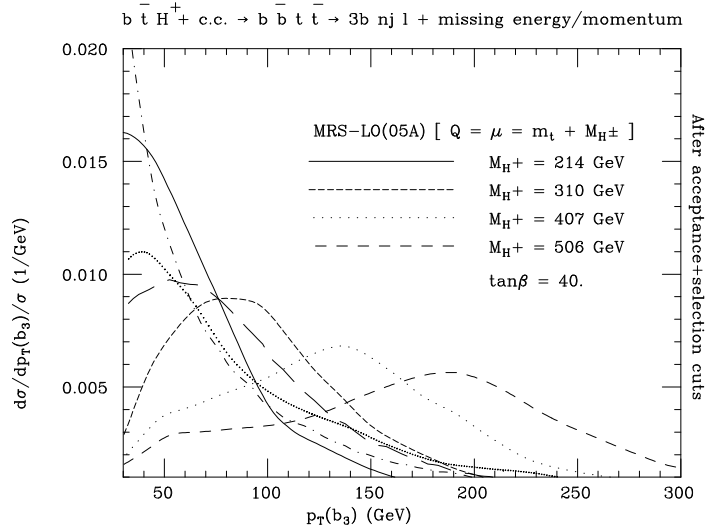}}
\caption{\small Differential distributions in transverse momentum of
the $b$-quark accompanying the reconstructed $t\bar t$ pair 
in the signal (2)--(3), 
for four selected values of $M_{H^\pm}$ in the heavy mass range,
with $\tan\beta=40$, after the acceptance 
and selection cuts described in the text: i.e. eqs.~(15)--(16)
and steps (a)--(d). 
The PDF set used was MRS-LO(05A)
with renormalisation and factorisation scales set equal 
to $m_t+M_{H^\pm}$. 
The (fine-dotted)[long-dashed]\{dot-dashed\} curve 
represents the shape of the background process
((10))[(11)]\{(12)\}. Normalisation is to unity.}
\label{fig:pT}
\end{center}
\end{figure}

Fig. 2 shows the $p_T$ distribution of the 3rd $b$-jet accompanying
the reconstructed $t\bar t$ pair as discussed above in step (e).  We
clearly see a harder $p_T$ distribution for the signal compared to the
background processes for a $H^\pm ~{\rm mass} ~ \geq 300$ GeV.  Thus we
can improve  the signal/background ratio over this mass range by imposing a 
\be
p_T > 80 ~{\rm GeV}
\label{seventeen}
\ee
cut on this 3rd $b$-jet.

\begin{figure}[!t]
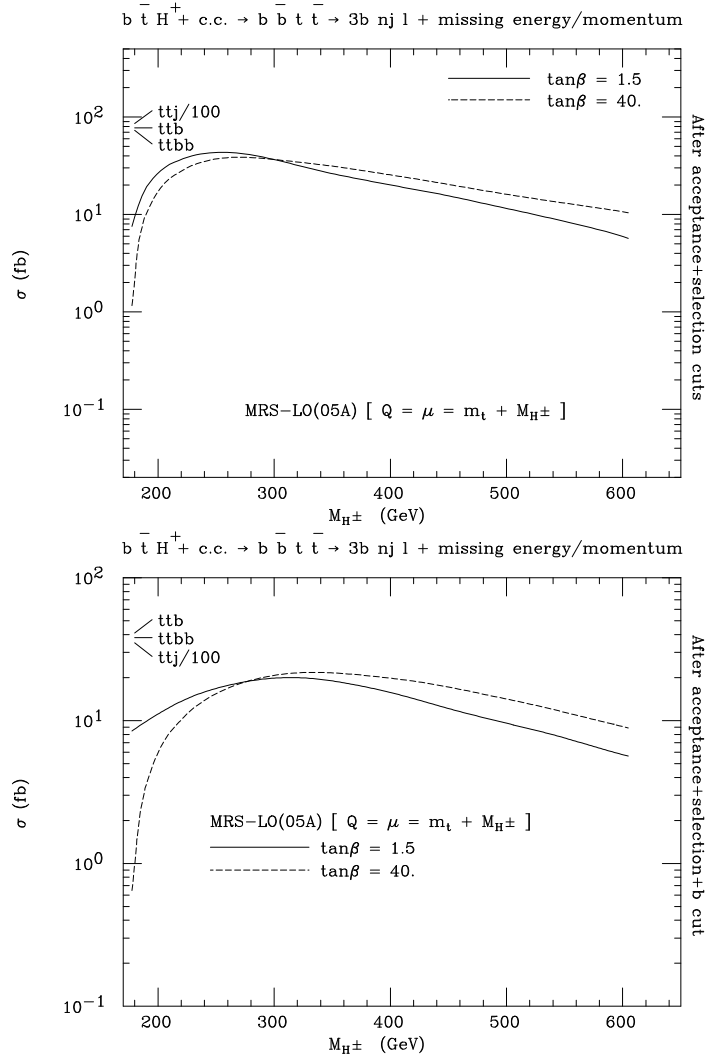

\begin{center}
{\epsfig{file=newselect.ps,height=9.25cm,angle=90}}
{\epsfig{file=newfinal.ps,height=9.25cm,angle=90}}
\caption{\small Production cross section for the signal  (2)--(3)
as a function of $M_{H^\pm}$  in the heavy mass range,
 after the acceptance 
and selection cuts described in the text: i.e.  eqs.~(15)--(16)
and steps (a)--(d) (top plot) as well as the transverse momentum cut
(17) on the $b$-jet accompanying the top-antitop pair (bottom plot). 
The PDF set used was MRS-LO(05A)
with renormalisation and factorisation scales set equal \
to $m_t+M_{H^\pm}$. 
The arrows represent the size of the backgrounds 
(10)--(12), the last of
which has been divided by 100 (for readability).
No $b$-tagging efficiency/rejection is included.}
\label{fig:final}
\end{center}
\end{figure}

The top of Fig. 3 shows the signal cross-section along with those of the
background processes (10)--(12) after applying the selection cuts
(15)--(16) and the mass constraints of steps (a)--(d).  No $b$-tagging
efficiency or rejection factor has been applied yet.  The effect of
imposing the $p_T$-cut (17) on the 3rd $b$-jet is presented in the bottom
plot.  It is clearly shown to suppress the backgrounds significantly:
in particular the dominant one from $ttj$ (12) is reduced by a factor
of 2.5 or so.  In contrast the signal cross-section is essentially
unaffected for a $H^\pm$ mass $\geq 400$ GeV.

Finally, Fig. 4 shows the signal and background cross-sections against
the reconstructed $bt$ invariant mass as discussed in step (f).
Here we have included a $b$-tagging efficiency of 40\% and a probability
of 1\% for mis-tagging a light quark or gluon jet $(j)$ as $b$-jet [11].
The top figure shows the signal and background cross-sections
separately while the bottom one shows their sum for different $H^\pm$
masses at $\tan\beta = 40$.  The signal peaks are clearly visible in
the latter.  One gets similar results for $\tan\beta = 1.5$.

\begin{figure}[!h]
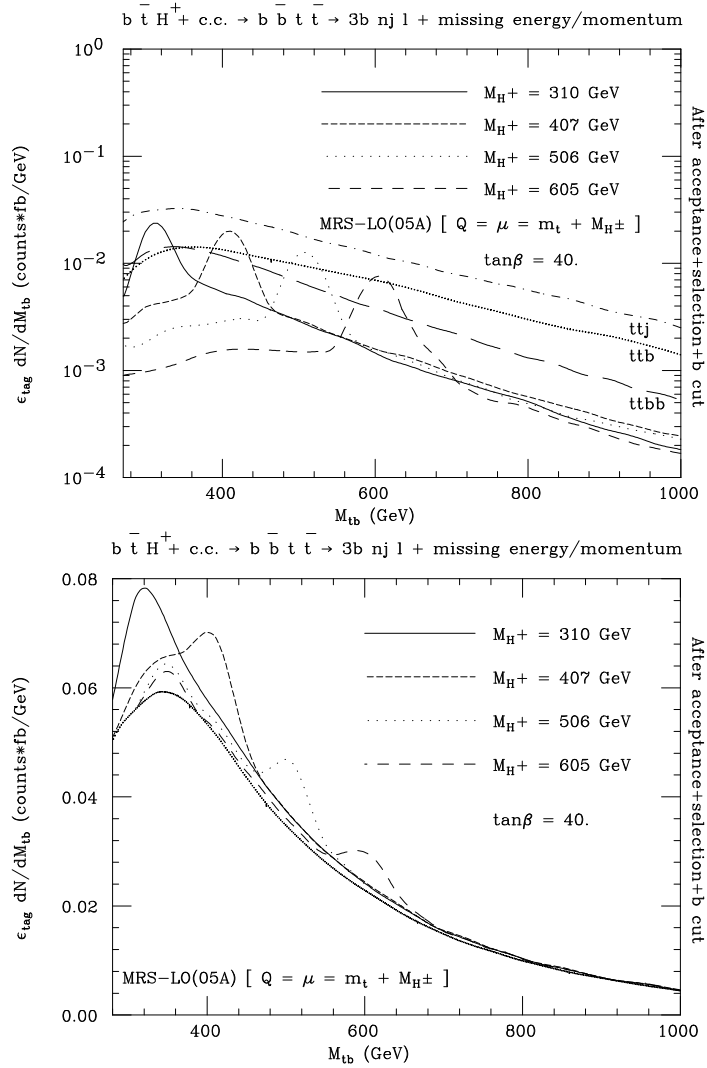

\begin{center}
{\epsfig{file=newreso2.ps,height=9.25cm,angle=90}}
{\epsfig{file=newheavy.ps,height=9.25cm,angle=90}}
\caption{\small (Top plot) Differential distribution (two entries per
each event generated) in the reconstructed charged
Higgs mass for the signal (2)--(3), corresponding to
four selected values of $M_{H^\pm}$ in the heavy mass range,
for $\tan\beta=40$, after the acceptance 
and selection cuts described in the text: i.e. eqs.~(15)--(16)
and steps (a)--(d) as well as the transverse momentum cut
(17) on the $b$-jet accompanying the top-antitop pair. 
The PDF set used was MRS-LO(05A)
with renormalisation and factorisation scales set equal 
to $m_t+M_{H^\pm}$. 
The (fine-dotted)[long-dashed]\{dot-dashed\} curve 
represents the shape of the background process
(({10}))[({11})]\{(12)\}.
(Bottom plot) As above, after summing each signal to all 
 backgrounds.
Tagging efficiencies have been included here, in both plots.}
\label{fig:reso}
\end{center}
\end{figure}

Tab. 1 lists the number of signal and background events over a 80 GeV
bin around the $H^\pm$ mass for an annual luminosity of 100
fb$^{-1}$, expected from the high luminosity run of the LHC.  The
corresponding values of $S/\sqrt{B}$ are also shown.  We see a better
than 5(3)$\sigma$ signal up to a $H^\pm$ mass of 400(600) GeV at
$\tan\beta = 40$.  It may be noted that both the signal size and the
$S/\sqrt{B}$ ratio are better here in comparison with the 4 $b$-tagged
channel [4] for $\epsilon_b = 40\%$ and $p_T$ cut of 30 GeV.  But the
$S/B$ ratio is better in the latter case.  Similar results hold for
$\tan\beta = 1.5$.

\begin{table}[!h]
\begin{center}
\begin{tabular}{|c||c|c|c|}
\hline
\multicolumn{4}{|c|}
{Number of events per year}\\
\hline
$M_{H^\pm}\pm40$ GeV & $S$ & $B$ & $S/\sqrt B$ \\
\hline
$310$ & 
$133$  & 
$443$ & 
$6.2$ \\
$407$ & 
$111$  & 
$403$ & 
$5.6$ \\
$506$ & 
$73$  & 
$266$ & 
$4.5$ \\
$605$ & 
$43$  & 
$156$ & 
$3.4$ \\
\hline\hline
\multicolumn{4}{|c|}
{MRS-LO [~$Q=\mu=m_t+M_{H^\pm}$~]} \\
\hline\hline
\multicolumn{4}{|c|}
{$3b~+~n~{\mathrm{jets}}~+~\ell^\pm~+~p^T_{\mathrm{miss}}$
(with $n=2,3$)\qquad \qquad 
After all cuts}
\\ \hline
\end{tabular}
\end{center}
\caption{\small Number of events from  the signal (2)--(3),
$S$, and the sum of the backgrounds 
(10)--(12), $B$, along with the
statistical significance, ${\protect{S/\sqrt B}}$, 
per 100 inverse femtobarns of integrated luminosity, in a window of
80~GeV around four selected values of
 $M_{H^\pm}$ (given in GeV) in the heavy mass range, 
for $\tan\beta=40$. At least three $b$-jets are assumed to be tagged,
each with efficiency $\epsilon_b=40\%$, whereas the rejection factor
against light jets is $\epsilon_{j=q,g}=1\%$.
All cuts discussed in the text,  i.e.  eqs.~(15)--(16),
steps (a)--(d) as well as the transverse momentum cut
(17) on the $b$-jet accompanying the top-antitop pair, 
have been enforced.  
The PDF set used was MRS-LO(05A)
with renormalisation and factorisation scales set equal 
to $m_t+M_{H^\pm}$.}
\label{tab:discovery}
\end{table}

In summary, the isolated lepton + multi-jet channel with triple
$b$-tagging -- supplemented by a transverse momentum cut on the third $b$-jet
-- offers a promising signature for $H^\pm$ searches at the LHC
up to $M_{H^\pm}\approx 600$ GeV 
at $\tan\beta \gsim 40$ and $\lsim 1.5$, thus extending the
reach of previous similar analyses [2,3].  Hence it calls
for a more detailed study, including hadronisation, jet
identification and detector effects.
\bigskip

\noindent {\bf Acknowledgements:} {\small 
This work was started at the Les Houches
Workshop on Physics at TeV Colliders, organised by LAPP, Annecy.  We
thank the organisers, Patrick Aurenche and Fawzi Boudjema, for a very
stimulating environment.  The work of DPR was partly supported by the
IFCPAR under project No. 1701-1. SM acknowledges financial support from
the UK-PPARC.}

\vfill\newpage
\noindent {\Large\bf References}

\begin{enumerate}
\item[{[1]}] J.F. Gunion, H.E. Haber, G.L. Kane and S. Dawson, ``The
Higgs Hunters' Guide'' (Addison-Wesley, Reading, MA, 1990).
\item[{[2]}] V. Barger, R.J.N. Phillips and D.P. Roy, Phys. Lett. B324
(1994) 236. 
\item[{[3]}] J.F. Gunion, Phys. Lett. B322 (1994) 125. 
\item[{[4]}] D.J. Miller, S. Moretti, D.P. Roy and W.J. Stirling,
hep-ph/9906230. 
\item[{[5]}] K. Odagiri, hep-ph/9901432; D.P. Roy, Phys. Lett. B459
(1999) 607. 
\item[{[6]}] A.A. Barrientos Bendez\'u and B.A. Kniehl, Phys. Rev. D59
(1999) 015009; S. Moretti and K. Odagiri, Phys. Rev. D59 (1999)
055008. 
\item[{[7]}] A.A. Barrientos Bendez\'u and B.A. Kniehl, hep-ph/9908385. 
\item[{[8]}] M. Drees, M. Guchait and D.P. Roy, hep-ph/9909266;
S. Moretti and K. Odagiri, in preparation. 
\item[{[9]}] F. Borzumati, J.-L. Kneur and N. Polonsky,
hep-ph/9905443. 
\item[{[10]}] D. Dicus, T. Stelzer, Z. Sullivan and S. Willenbrock,
Phys. Rev. D59 (1999) 094016. 
\item[{[11]}] E. Richter-Was and M. Sapinski, ATLAS note
(ATL-PHYS-98-132); V. Drollinger, T. Mueller and R. Kinnunen, CMS note
(1999/001). 
\item[{[12]}] A.D. Martin, R.G. Roberts, W.J. Stirling and
R.S. Thorne, Phys. Lett. B443 (1998) 301. 
\item[{[13]}] H.L. Lai et al., Phys. Rev. D55 (1997) 1280.
\item[{[14]}] R. Barnett, H. Haber and D. Soper, Nucl. Phys. B306
(1988) 697; F. Olness and W.K. Tung, Nucl. Phys. B308 (1988) 813;
M. Aivazis, J. Collins, F. Olness and W.-K. Tung, Phys. Rev. D50
(1994) 3085; Phys. Rev. D50 (1994) 3102. 
\item[{[15]}] A.D. Martin, R.G. Roberts, M.G. Ryskin and
W.J. Stirling, Eur. Phys. J. C2 (1998) 287; H.L. Lai and W.K. Tung,
Z. Phys. C74 (1997) 463; M. Buza, Y. Matiounine, J. Smith and W.L. van
Neerven, Eur. Phys. J. C1 (1998) 301; Phys. Lett. B411 (1997) 211. 
\item[{[16]}] L.G. Jin, C.S. Li, R.J. Oakes and S.H. Zhu,
hep-ph/9907482. 
\item[{[17]}] J.A. Coarasa, D. Garcia, J. Guasch, R.A. Jim\'enez
and J. Sol\`a, Eur. Phys. J. C2 (1998) 373.
\item[{[18]}] T. Stelzer and W.F. Long, Comp. Phys. Comm. 81, 357
(1994). 
\item[{[19]}] H. Murayama, I. Watanabe and K. Hagiwara, HELAS:
HELicity Amplitude Subroutines for Feynman Diagram Evaluations, {\sl
KEK Report} 91-11, January 1992.
\end{enumerate}

\end{document}